\documentclass[12pt]{iopart}
\usepackage{amssymb}
\usepackage{graphicx,bm}
\usepackage{setstack,cite}
\begin{document}

\title[Nonlinear coherent states]
{On the generalized intelligent states and certain
related nonclassical states of a quantum exactly solvable nonlinear oscillator
}

\author{V~Chithiika Ruby and  M~Senthilvelan$^{\dagger}$}

\address{Centre for Nonlinear Dynamics, School of Physics,
Bharathidasan University, Tiruchirapalli - 620 024, India.}
\ead{velan@cnld.bdu.ac.in}

\begin{abstract}
We construct nonlinear coherent states or f-deformed coherent states for a nonpolynomial nonlinear oscillator
which can be considered as placed in the middle between the harmonic oscillator and the isotonic oscillator
(Cari\~{n}ena J F et al, \emph{J. Phys. A: Math. Theor.} {\bf 41}, 085301 (2008)).
The deformed annihilation and creation operators which are required to construct the
nonlinear coherent states in the number basis are obtained from the solution of the Schr\"{o}dinger equation.
Using these operators, we construct generalized intelligent states,
nonlinear coherent states, Gazeau-Klauder coherent states and the even and odd nonlinear coherent states
for this newly solvable system.  We also report certain nonclassical properties exhibited by these nonlinear coherent states.
In addition to the above, we consider position dependent mass Schr\"{o}dinger equation associated with this solvable nonlinear
oscillator and construct nonlinear coherent states, Gazeau-Klauder coherent states and the even and odd
nonlinear coherent states for it. We also give explicit expressions of all these nonlinear coherent states by
considering a mass profile which is often used for studying transport properties in semiconductors.
\end{abstract}

\maketitle
\section{Introduction}
\label{sec1}
In a recent paper Cari\~{n}ena et al \cite{Carinena} have considered the potential,
\begin{eqnarray}
V_{g}(x) = \frac{1}{2}\left(\omega^2 x^2  +\frac{2g_a(x^2 - a^2)}{(x^2 + a^2)^2}\right), \quad g_a > 0,
\label{a1}
\end{eqnarray}
which can be considered as placed in the middle between the harmonic oscillator and the isotonic
oscillator potentials and shown that in a particular case, namely $g_a = 2\omega a^2(1+2\omega a^2)$,
$a^2 = \frac{1}{2}$ and $\omega = 1$, the associated Schr\"{o}dinger equation,
\begin{eqnarray}
\frac{d^2\psi(x)}{dx^2}+\frac{m_0}{\hbar^2}\left( 2E - x^2  -\frac{8(2x^2 - 1)}{(2x^2 +1)^2}\right)\psi(x) = 0
\label{a2}
\end{eqnarray}
is exactly solvable. The authors have obtained the eigenfunctions in terms of
${\cal P}$-Hermite functions (defined in Eq. (\ref{ap2})), namely
\begin{eqnarray}
\psi_n(x) &=& N_n \displaystyle{\frac{{\cal P}_n(x)}{(1+2x^2)}} e^{\displaystyle -\frac{x^2}{2}}, \quad n = 0,3,4,...,
\label{a3}
\end{eqnarray}
where the normalization constant is given by
\begin{eqnarray}
N_n =\left[\displaystyle{\frac{(n-1)(n-2)}{2^n n! \sqrt{\pi}}}\right]^{1/2}, \quad n = 0,3,4,....
\label{a4}
\end{eqnarray}
It is noted that while solving the Schr\"{o}dinger equation (\ref{a2}) the constants ($\hbar, m_0$) have
been taken as one for the sake of simplicity. The energy spectrum is given by
\begin{eqnarray}
 E_n &=& -\frac{3}{2} + n
\label{a5}
\end{eqnarray}
which shows that the ground state ($\psi_0$) has an energy $E_0$ which
is lower than that of the pure harmonic oscillator case  \cite{Carinena}.
Consequently the above quantum exactly solvable potential was analyzed in
different perspectives. In the following, we briefly summarize the activities
revolving around this inverse square potential.

To start with Fellows and Smith have shown that the solvable potential considered by
Cari\~{n}ena et al is a supersymmetric partner potential of the harmonic oscillator \cite{Fellow}.
In a different study, Kraenkel and Senthilvelan, have considered
the exactly solvable potential given by Cari\~{n}ena, and obtained a different class of exactly solvable potentials
by transforming the Schr\"{o}dinger equation (\ref{a2}) into
the position dependent mass Schr\"{o}dinger equation and solving the underlying equation.
They have also given bound state energies and their corresponding wavefunctions for the potentials
they have considered  \cite{Sen}.

The coherent states \cite{Gbook} for the position dependent mass Schr\"{o}dinger
equation was then constructed recently by the present authors \cite{chithiika}
with an illustration of the above exactly solvable nonlinear oscillator.
In a very recent paper, Sesma, has considered the
Schr\"{o}dinger equation associated with the potential (\ref{a1}) and transformed it into a
confluent Heun equation and solved it numerically \cite{Sesma}. For certain specific values of the parameters
the author has also constructed quasi-polynomial solutions.

The aim of this paper is to make some progress in constructing nonlinear coherent states
for the Schr\"{o}dinger Eq. (\ref{a2}) and bring out their
statistical properties \cite{Vbook}. We also construct nonlinear coherent states for the position
dependent mass Schr\"{o}dinger equation associated with this nonlinear oscillator.
Nonlinear coherent states are defined as the eigenstates of an operator $\hat{a}f(\hat{n})$,
which satisfy the eigenvalue equation $\hat{a}f(\hat{n})|\alpha,f\rangle = \alpha |\alpha,f\rangle$,
where $f(\hat{n})$ is an operator-valued function of the number
operator, $\hat{n} = \hat{a}^{\dagger}\hat{a}$, and they are nonclassical \cite{Nieto, Manko1}.
Nonlinear coherent states were first introduced explicitly in de Matos Filho and Vogel \cite{Filho}
and Man'ko et al \cite{Manko} but before them they were implicitly defined by Shanta et al \cite{Shan} in a
compact form (see also Ref. \cite{Rokin}). In the consequent years these states have been studied in depth and
shown to exhibit several nonclassical properties including squeezing, amplitude-squared squeezing, higher order squeezing ,
anti-bunching, sub-Poissonian statistics and oscillatory number distribution \cite{Walls, squee, Paul, Mah, Dodo1}.
A veritable explosion of activities occurred in the past two decades in the study of
nonlinear coherent states and their underlying properties \cite{Vogel, Manko2}.

It is known that once suitable deformed ladder operators are found then the nonlinear coherent states can be constructed in a straightforward manner
for the given quantum system.  As we pointed earlier the authors in \cite{Fellow} have obtained the ladder operators for (\ref{a2})
through the supersymmetric technique and shown that the operators are related to the
intertwining between the Hamiltonian (\ref{a2}) and the harmonic oscillator.  In this work we build the deformed
ladder operators from the solution of the Schr\"{o}dinger equation \cite{Dong, Ahmed}
and show that these operators satisfy the necessary requirements,
 \begin{eqnarray}
\hat{A}|n\rangle \propto |n-1\rangle, \nonumber \\
\hat{A}^{\dagger}|n\rangle \propto |n+1\rangle,
\label{f2}
\end{eqnarray}
to construct the nonlinear coherent states in
Fock space representation \cite{Barut, Gazeau, Klauder}.  To construct these operators we
derive two new recurrence relations solely in terms of the
${\cal P}$-Hermite polynomials (vide Eqs. (\ref{ap10}) and (\ref{ap15})).  We then rewrite these
two recurrence relations suitably and extract the necessary creation and annihilation operators.

Interestingly, we find that the
intensity dependent function $f(n)$ has zeroes at $n=1$ and $3$.
Hence, the Fock space breaks up into a countable number of irreducible representations \cite{Manko}.
These reduced pieces will not allow a unitary representation \cite{Glau}. System
(\ref{a2}) may be considered as an example for the theory proposed in Refs. \cite{Glau, Manko}.

From the deformed ladder operators, we construct generalized intelligent states. To do so
we introduce two Hermitian operators, say $\hat{W}$ and $\hat{P}$ in
terms of deformed annihilation and creation operators, related to the uncertainty relation
\cite{Schr,Robert} $(\Delta \hat{W})^2 (\Delta \hat{P})^2 \ge \frac{1}{2}\langle G \rangle^2$
with $\hat{G} = i[\hat{W}, \hat{P}]$. The states which satisfy the equality relation in
the above uncertainty relation are called the intelligent states \cite{Arag, Dodo, Kinani}.
They also represent the squeezed states \cite{Walls}.
We then construct nonlinear coherent states by taking $\lambda = 1$ in the generalized intelligent states.
We also study certain Hilbert space properties and nonclassical properties
for this quantum system. From the deformed Hamiltonian corresponding to
this function $e_{n} = nf^{2}(n)$, we construct the Gazeau-Klauder coherent states which are defined
as coherent states that also saturate the Heisenberg-uncertainty relation. We also
construct even and odd nonlinear coherent states \cite{Man, Siva}.
Motivated from the recent developments in the study of the position dependent mass Schr\"{o}dinger equation,
we also construct nonlinear coherent states for the exactly solvable
position dependent mass Schr\"{o}dinger equation associated with the Eq. (\ref{a2}).
Finally, we consider a specific mass profile which is quite often used to study transport properties in semiconductors and give explicit expressions for all the 
coherent states we have considered \cite{Koc,Miller}. The results given in this paper
give further insights into the geometrical properties exhibited by this system.

This paper is organized as follows.  In Sec 2, we discuss the method of finding
deformed annihilation and creation operators in the number basis for the nonlinear oscillator.
In Sec 3, we construct the generalized intelligent states for this nonlinear oscillator.  In Sec 4, we construct
the nonlinear coherent states for this oscillator and show that the states satisfy completeness condition.
We then deduce certain statistical properties associated with the nonlinear coherent states.
In Sec 5, we construct Gazeau-Klauder coherent states for this nonlinear oscillator and bring out their
statistical properties. In Sec 6, we derive even and odd nonlinear coherent states for this system.
In Sec 7, we consider the position dependent mass Schr\"{o}dinger equation associated with this
inverse type potential and construct the nonlinear coherent states for it. We also illustrate the theory with an
example. Finally we present our conclusions in Sec. 8. In Appendix, we present the derivation of two
recurrence relations which are essential to construct the ladder operator for this system (\ref{a2}).

\section{Ladder operators}
\label{sec2}
We consider Eq. (\ref{a2}) as the number operator equation after subtracting the ground state energy $E_0 = -\frac{3}{2}$
\begin{eqnarray}
\hat{n}|n\rangle = n|n\rangle
\label{a6}
\end{eqnarray}
and address the problem of finding deformed creation ($\hat{A}^{\dagger}$) and
annihilation operators ($\hat{A}$) for the given wave function  \cite{Dong}.

To begin with recall the two recurrence relations, (\ref{ap10}) and (\ref{ap15}).  Multiplying
each of them by $N_ne^{-\frac{x^2}{2}}/(1+2x^2)$ and substituting (\ref{a3}) and its
derivative in them and rewriting the resultant expressions suitably one can reexpress the recurrence
relations in terms of the wavefunction $\psi_n$.  The resultant expressions read
\begin{eqnarray}
\fl \;\; \left[n-1 + \frac{2(2x^2-1)}{(1+2x^2)^2}\right]\psi^{'}_n + \left[nx-\phi+\frac{2(2x^2-1)}{(1+2x^2)^2}\phi\right]
\psi_n = \sqrt{2n(n-1)(n-3)} \psi_{n-1},
\label{an1}
\end{eqnarray}
\begin{eqnarray}
\fl \;\;\; -\left[n + \frac{2(2x^2-1)}{(1+2x^2)^2}\right]\psi^{'}_n + \left[nx-\frac{2(2x^2-1)}{(1+2x^2)^2}\phi\right]
\psi_n = \sqrt{2(n+1)n(n-2)} \psi_{n+1},
\label{an2}
\end{eqnarray}
where prime denotes differentiation with respect to $x$.

From these expressions one can extract the deformed annihilation, $\hat{A}$ and
creation operators, ($\hat{A}^{\dagger}$) of the form
\begin{eqnarray}
\sqrt{2}\hat{A}=\left[\frac{2(2x^2 - 1)}{(1+2x^2)^2}-1\right]\left[\frac{d}{dx} + \phi\right]+\left[\frac{d}{dx} + x \right]\hat{n},
\label{a13}
\end{eqnarray}
\begin{eqnarray}
\sqrt{2}\hat{A}^{\dagger}=-\frac{2(2x^2 - 1)}{(1+2x^2)^2}\left[\frac{d}{dx} + \phi\right]+\left[-\frac{d}{dx} + x \right]\hat{n}.
\label{a17}
\end{eqnarray}

These deformed ladder operators $\hat{A}$ and $\hat{A}^{\dagger}$ satisfy the relations
\begin{eqnarray}
\hat{A}|n\rangle &=& \sqrt{n}\;f(n)\;|n-1\rangle, \\
\hat{A}^{\dagger}|n\rangle &=& \sqrt{n+1}\;f(n+1)\; |n+1\rangle,
\label{a18}
\end{eqnarray}
with $f(n) = \sqrt{(n-1)(n-3)}$. Hence, the creation and annihilation operators in the Fock space
for the Schr\"{o}dinger equation (\ref{a6}) can be written as
\begin{eqnarray}
\hat{a} = \hat{A}\frac{1}{f(\hat{n})}, \qquad \hat{a}^{\dagger} = \hat{A}^{\dagger}\frac{1}{f(\hat{n}+1)}
\label{a19}
\end{eqnarray}
which in turn satisfy the following relations
\begin{eqnarray}
\hat{a}|n\rangle &=& \sqrt{n}\;|n-1\rangle, \nonumber \\
\hat{a}^{\dagger}|n\rangle &=& \sqrt{n+1}\;|n+1\rangle.
\label{a20}
\end{eqnarray}
Thus we obtain the Heisenberg algebra in the number basis $|n\rangle$ through the set of
the elements $(\hat{a},\hat{a}^{\dagger}$ and $I)$ satisfying the identities
\begin{eqnarray}
[\hat{a},\hat{a}^{\dagger}]|n\rangle = |n\rangle, \qquad [\hat{a},\hat{n}]|n\rangle = -\hat{a} |n\rangle,
\qquad [\hat{a}^{\dagger},\hat{n}]|n\rangle = \hat{a}^{\dagger}|n\rangle,
\label{a21}
\end{eqnarray}
where $\hat{n} = \hat{a}^{\dagger}\hat{a}$.

\section{Generalized intelligent states}
\label{sec3}
We define two Hermitian operators, $\hat{W}$ and $\hat{P}$, in terms of the deformed creation and annihilation operators, $\hat{A}^{\dagger}$ and
 $\hat{A}$, \cite{Kinani}:
\begin{eqnarray}
\hat{W} = \frac{1}{\sqrt{2}}(\hat{A}+ \hat{A}^{\dagger}), \qquad \hat{P} = \frac{i}{\sqrt{2}}(\hat{A}^{\dagger}- \hat{A}),
\label{b1}
\end{eqnarray}
which also satisfy the commutation relation $[\hat{W}, \hat{P}] = i\hat{G}$.
It has been shown that the Hermitian operators $\hat{W}$ and $\hat{P}$ satisfying this non-canonical commutation relation,
the variances, $(\Delta \hat{W})^2$ and $(\Delta \hat{P})^2$, satisfy the Heisenberg uncertainty relation \cite{Schr} $
(\Delta \hat{W})^2 (\Delta \hat{P})^2 \ge \frac{1}{4} \langle \hat{G} \rangle^2$.
The generalized intelligent states are obtained when the equality in the Heisenberg uncertainty relation
is realized \cite{Dodo}. The generalized intelligent states satisfy the eigenvalue equation
\begin{eqnarray}
(\hat{W}+i\lambda \hat{P})|\psi\rangle = \sqrt{2} \alpha |\psi\rangle, \quad \lambda, \alpha \in \mathbb{C}.
\label{b4}
\end{eqnarray}
It is proved that the parameter $\lambda$ in (\ref{b4}) is determined by
$|\lambda| = \frac{(\Delta\hat{W})}{(\Delta\hat{P})}$ and can be interpreted as the
control parameter for the squeezing effect in the states $|\psi\rangle$.

To solve the eigenvalue Eq. (\ref{b4}), we invoke the definition (\ref{b1})
so that the eigenvalue equation (\ref{b4}) becomes
\begin{eqnarray}
\left[(1-\lambda) \hat{A}^{\dagger} + (1+\lambda)\hat{A}\right]|\psi\rangle = 2\alpha|\psi\rangle.
\label{b5}
\end{eqnarray}
We assume $|\psi\rangle$ be of the form
\begin{eqnarray}
|\psi\rangle = |\alpha, f, \lambda\rangle = \sum^{\infty}_{n=0} c_n |n\rangle, \qquad n \neq 1,2,
\label{b7}
\end{eqnarray}
where the coefficients, $c_{n}\;'s,\;n=0,3,4,5,..$, determined by (\ref{b5})
. Substituting Eq. (\ref{b7}) in (\ref{b5}), we get
\begin{eqnarray}
(1-\lambda) \sum^{\infty}_{n=0} c_n \hat{A}^{\dagger}|n\rangle + (1+\lambda)\sum^{\infty}_{n=0}c_n \hat{A}|n\rangle = 2\alpha \sum^{\infty}_{n=0}c_n|n\rangle.
\label{b8}
\end{eqnarray}
Equating the coefficients of $\psi_n$ in the above Eq. (\ref{b8}) yields
\begin{eqnarray}
\fl \qquad (1-\lambda)\sqrt{n(n-1)(n-3)}\;c_{n-1}+(1+\lambda)\sqrt{(n+1)n(n-2)}\;c_{n+1} = 2 \alpha c_n.
\label{b9}
\end{eqnarray}

From (\ref{b9}), we find $c_0 = 0$. To obtain the value of other coefficients $c_i\;'s,\;i = 3,4,5,\ldots,n$,
we rewrite Eq. (\ref{b9}) of the form
\begin{eqnarray}
\frac{(1-\lambda)}{(1+\lambda)}\sqrt{n}\;f(n)\frac{c_{n-1}}{c_n}+ \sqrt{n+1}\;f(n+1)\frac{c_{n+1}}{c_n} = \frac{2\alpha}{(1+\lambda)}.
\label{b10}
\end{eqnarray}
Defining $B_n = \frac{c_{n+1}}{c_n}$ and $B_{n-1} = \frac{c_{n}}{c_{n-1}}$, Eq. (\ref{b10})
can be brought to the form
\begin{eqnarray}
B_{n} = \frac{1}{\sqrt{n+1}\;f(n+1)}\left[\frac{2\alpha}{(1+\lambda)}+\left(\frac{\lambda-1}{\lambda+1}\right)\frac{n f^{2}(n)}{B_{n-1}}\right].
\label{b11}
\end{eqnarray}
Eq. (\ref{b11}) can be reutilized to express $B_{n-1}$ in terms of $B_{n-2}$ and $B_{n-2}$ in terms of
$B_{n-3}$ and so on.  Substituting all these expressions in (\ref{b11}) one arrives at
\begin{equation}
\fl B_n= \frac{1}{\sqrt{n+1}f(n+1)}\Biggl[\frac{2\alpha}{1+\lambda }+\frac{ \left(\frac{\lambda -1}{\lambda +1}%
\right) nf^{2}(n)}{\frac{2\alpha}{1+\lambda }+\frac{\left( \frac{\lambda -1}{%
\lambda +1}\right) (n-1)f^{2}(n-1)}{ \frac{2\alpha}{1+\lambda }+\frac{\left( \frac{%
\lambda -1}{\lambda +1}\right) (n-2)f^{2}(n-2)}{%
\begin{array}{c}
\hspace{-5.5cm}\frac{2\alpha}{1+\lambda }+.... \\ \hspace{-2.5cm}+.....
\\ \hspace{3.0cm}  \frac{2\alpha}{1+\lambda }+\frac{ \left( \frac{\lambda
-1}{\lambda +1}\right) (4)f^{2}(4)}{ \frac{2\alpha}{1+\lambda}}\Biggr].
\end{array}
}}} \label{b20}
\end{equation}
Thus starting from the definition $c_n=B_{n-1}c_{n-1}$ one finds that all the coefficients
in the series, (\ref{b7}), $c_n 's,\; n=4,5,6,\ldots$, can be expressed in terms of $c_3$, that is
\begin{eqnarray}
c_n = B_{n-1} B_{n-2} B_{n-3} ... B_3 c_3 = \tilde{B}_{n-1}!c_3, \quad n = 3,4,5,,...,
\label{b15}
\end{eqnarray}
with $\tilde{B}_2! = 1$.  As a consequence one
can express $|\psi \rangle$ of the form
\begin{eqnarray}
|\alpha, \tilde{f}, \lambda\rangle = c_3 \sum^{\infty}_{n=3} \tilde{B}_{n-1}!|n\rangle .
\label{b22}
\end{eqnarray}

In order to compare (\ref{b22}) with the latter expressions given in the paper we redefine the
constants, $\tilde{B}_{n-1}!, \quad n = 3,4,5,\ldots$, as
\begin{eqnarray}
\tilde{B}_{n-1}!= \frac{\tilde{A}_n !}{\sqrt{\tilde{n}!}\tilde{f}(n)!}
\label{b12}
\end{eqnarray}
where $\tilde{A}_n! = A_nA_{n-1}A_{n-2}...A_4$, $\tilde{f}(n)! = f(n)f(n-1)f(n-2)...f(4)$,
$\tilde{n}! = n(n-1)(n-2)...4$, and $\tilde{A}_3! = \tilde{f}(3)! = \tilde{3}! = 1$.
In the new variables. Eq. (\ref{b22}) read
\begin{eqnarray}
|\alpha, \tilde{f}, \lambda\rangle = c_3 \sum^{\infty}_{n=3}\frac{\tilde{A}_n !}{\sqrt{\tilde{n}!}\tilde{f}(n)!} |n\rangle .
\label{b23}
\end{eqnarray}
This is known as generalized intelligent states.  These are investigated with the value $|\lambda| = 1$ in the following section.
\section{Nonlinear coherent states}
\label{sec4}
In this section, we deduce nonlinear coherent states from the generalized intelligent states
by setting $\lambda = 1$ in the latter. The eigenvalue equation
for $\hat{A}$ with the eigenfunction $|\alpha, f\rangle$ in the Hilbert space reads
\begin{eqnarray}
\hat{A}|\alpha, f\rangle = \alpha |\alpha, f\rangle,  \qquad \alpha \in \mathbb{C}.
\label{c1}
\end{eqnarray}
Since $\lambda = 1$, we have $\Delta \hat{W} = \Delta \hat{P}$. Let
\begin{eqnarray}
|\alpha, f\rangle = \sum^{\infty}_{n=0} c_n |n\rangle, \qquad n \neq 1,2.
\label{c2}
\end{eqnarray}
Substituting Eq. (\ref{c2}) in Eq. (\ref{c1}), one gets
\begin{eqnarray}
\sum^{\infty}_{n=0} c_n \sqrt{n(n-1)(n-3)} |n-1\rangle = \alpha \sum^{\infty}_{n=0} c_n |n\rangle.
\label{c3}
\end{eqnarray}

Equating the coefficients of $\psi_n$ in the above Eq. (\ref{c3}) yields
\begin{eqnarray}
c_{n+1} \sqrt{n+1}\; f(n+1) = \alpha c_n
\label{c4}
\end{eqnarray}
from which we find $ c_0 = 0 $ and $c_3$ is an arbitrary constant. To express $c_n$ in terms of $c_3$
we rewrite the recurrence relation (\ref{c4}) in the form
\begin{eqnarray}
c_{n} = \frac{\alpha}{\sqrt{n}\;f(n)} c_{n-1}.
\label{c5}
\end{eqnarray}
From Eq. (\ref{c5}) one can find
\begin{eqnarray}
c_n = \frac{\alpha^{n-3}}{\sqrt{\tilde{n}!}\; \tilde{f}(n)!}c_3, \quad n = 3,4,5,....
\label{c9}
\end{eqnarray}
The coherent states defined for the oscillator (\ref{a2}) do not contain the states with photon number
less than three. The coherent states contain the states with photon number starting
from three reads now
\begin{eqnarray}
|\alpha, \tilde{f}\rangle = c_3 \sum^{\infty}_{n=3}\frac{\alpha^{n-3}}{\sqrt{\tilde{n}!}\;\tilde{f}(n)!}|n\rangle.
\label{c10}
\end{eqnarray}

To determine $c_3$, we use the normalization condition $\langle \alpha, \tilde{f}|\alpha,
\tilde{f}\rangle = 1$ which in turn gives
\begin{eqnarray}
c_3 = \left(\sum^{\infty}_{n=3}\frac{|\alpha|^{2n-6}}{\tilde{n}!\; [\tilde{f}(n)!]^2}\right)^{-1/2}.
\label{c12}
\end{eqnarray}
Since $c_3$ depends on $\alpha, \tilde{f}$, we denote it as $\tilde{N}$.
The nonlinear coherent states for the nonlinear oscillator Eq. (\ref{a6}) after appropriate
normalization reads
\begin{eqnarray}
|\alpha, \tilde{f}\rangle = \tilde{N}(|\alpha|^2)\sum^{\infty}_{n=3}\frac{\alpha^{n-3}}{\sqrt{\tilde{n}!}\;\tilde{f}(n)!}|n\rangle.
\label{c13}
\end{eqnarray}

\subsection{Completeness condition}
\label{ssec1}
In this subsection, we investigate whether the nonlinear coherent states form a complete system
of states in the Hilbert space or not.  To establish this we invoke
the completeness relation \cite{Klauder2, Chat}
\begin{eqnarray}
\frac{1}{\pi} \int \int_\mathbb{C} |\alpha, \tilde{f}\rangle W(|\alpha|^2)\langle \alpha, \tilde{f}| d^2\alpha = \hat{I},
\label{d1}
\end{eqnarray}
where $W(|\alpha|^2)$ is a positive weight function and $\hat{I}$ is an identity operator.
From (\ref{d1}) we obtain
\begin{eqnarray}
\frac{1}{\pi} \int \int_\mathbb{C} \langle\psi|\alpha, \tilde{f}\rangle W(|\alpha|^2)\langle \alpha, \tilde{f}|\Phi\rangle d^2\alpha = \langle\psi|\hat{I}|\Phi\rangle.
\label{d3}
\end{eqnarray}
Substituting $|\alpha,\tilde{f}\rangle$ and its conjugate (vide Eq. (\ref{c13})) in the left hand side of Eq. (\ref{d3}),
we get (which we call $G$)
\begin{eqnarray}
\fl \qquad G = \frac{1}{\pi} \sum^{\infty}_{m,n=3}\frac{\langle\psi|n\rangle \langle m|\Phi\rangle}{\sqrt{\tilde{n}!\tilde{m}!} \tilde{f}(n)!\tilde{f}(m)!}  \int \int_\mathbb{C} \alpha^{n-3}{\alpha^{*}}^{m-3} \tilde{N}^2(|\alpha|^2) W(|\alpha|^2)d^2\alpha.
\label{d4}
\end{eqnarray}

Taking $\alpha = re^{i\theta}$, one can separate the real and imaginary parts and obtain
\begin{eqnarray}
\fl \quad G = \frac{1}{\pi}\sum^{\infty}_{m,n=3}\frac{\langle\psi|n\rangle\langle m|\Phi\rangle}{\sqrt{\tilde{n}!\tilde{m}!}\; \tilde{f}(n)!\tilde{f}(m)!} \int^{\infty}_{0} \tilde{N}^{2}(r^2) r^{n+m-6}W(r^2)r dr \int^{2\pi}_{0} e^{i(n-m)\theta}d\theta.
\label{d5}
\end{eqnarray}
Since the second integral vanishes except $n = m$ we can bring Eq. (\ref{d5}) to the form
\begin{eqnarray}
G &=& \sum^{\infty}_{n=3}\frac{\langle\psi|n\rangle\langle n|\Phi\rangle}{\tilde{n}!\; [\tilde{f}(n)!]^{2}} \int^{\infty}_{0} \tilde{N}^{2}(r^2) r^{2n-6}W(r^2) 2 r dr.
\label{d6}
\end{eqnarray}
Taking $r^2 = x$ and using the identities $\tilde{n}! = \frac{n!}{6}$ and
$[\tilde{f}(n)!]^{2}= [(n-1)(n-3)]!$, we find
\begin{eqnarray}
G = \sum^{\infty}_{n=3}\frac{6\langle \psi|n\rangle\langle n|\Phi\rangle}{n![(n-1)(n-3)]!} \int^{\infty}_{0} x^{n-3}\tilde{N}^{2}(x)W(x)dx.
\label{d7}
\end{eqnarray}

Choosing $ \tilde{N}^{2}(x) W(x) = \frac{2}{3} x^{n^{2}-4n+7}K_{n^{2}-5n+3}(2x)$, the integral on the right hand side in
(\ref{d7}) can be brought to the form $\int^{\infty}_{0} x^{n^2 - 3n + 4} K_{n^2 - 5n + 3}(2x) dx$
which can be integrated in terms of gamma functions \cite{book}, namely $\frac{1}{4} \Gamma(n^{2}-4n+3) \Gamma(n+1)$.
As a result one gets
\begin{eqnarray}
G &=& \sum^{\infty}_{n=3}\langle\psi|n\rangle\langle n|\Phi \rangle.
\label{d9}
\end{eqnarray}
We mention here that the nonlinear coherent states contain the states with photon number greater than or equal to three.  As a consequence Eq. (\ref{d9}) turns out to be a projector for which the state $n = 0$ is excluded.

\subsection{Photon statistical properties of nonlinear coherent states $|\alpha,\tilde{f}\rangle$}
\label{ssec2}
Having constructed the nonlinear coherent states and studied its completeness,
in this subsection, we study certain statistical properties, namely, (1) the photon number distribution,
(2) sub-Poissonian function and (3) the correlation function, associated with these
nonlinear coherent states.

\begin{figure}[!ht]
\vspace{-0.5cm}
\begin{center}
\includegraphics[width=0.95\linewidth]{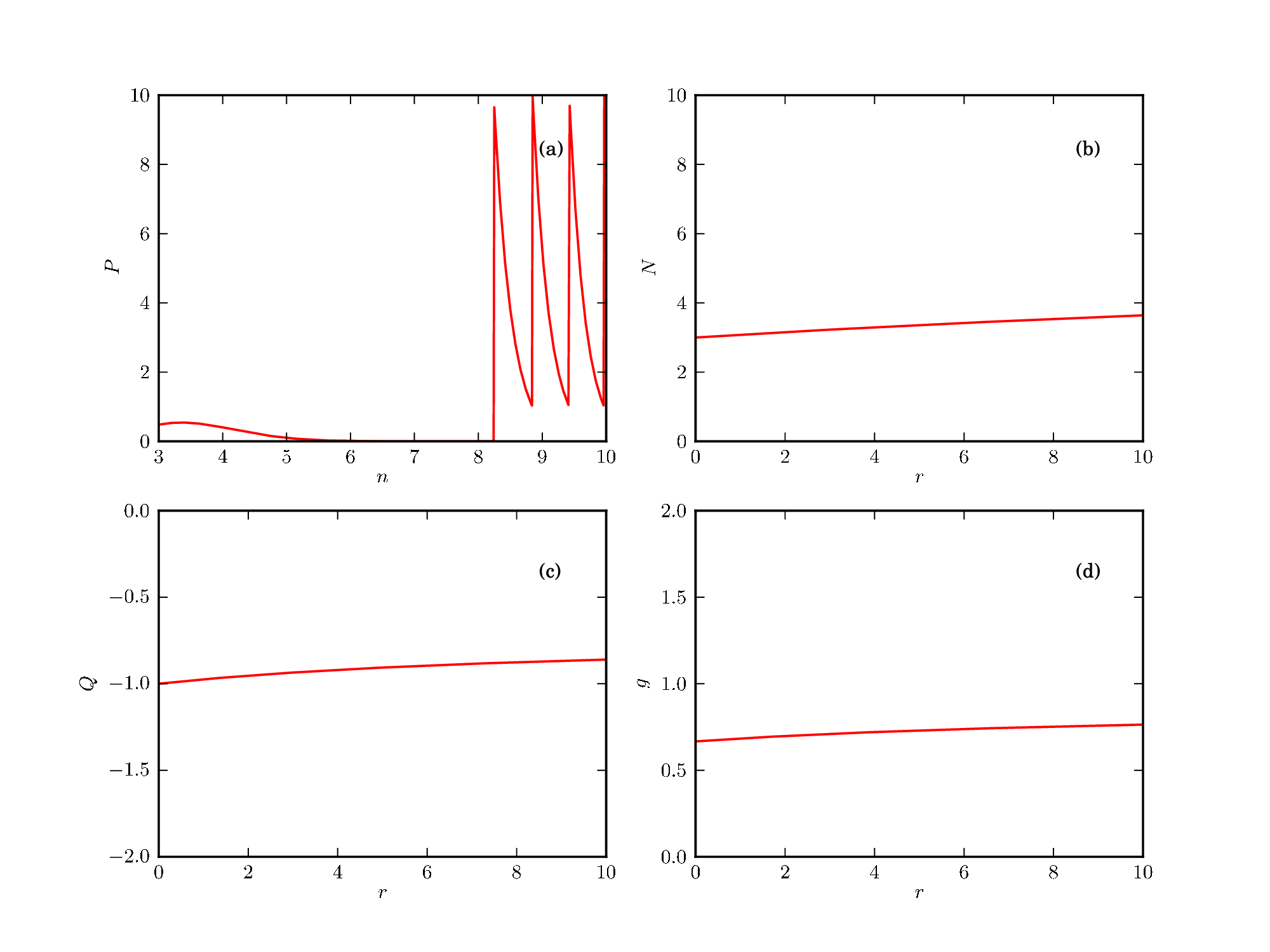}
\end{center}
\vspace{-1cm}
\caption{The plots of (a) photon number distribution $P(n)$, (b) mean photon number $\langle \hat{n} \rangle$ (denoted as $N$), (c) Mandel parameter $Q$  and (d) the second order correlation function $g^{2}(0)$ (denoted as $g$) of nonlinear coherent states (\ref{c13}).} \label{first}
\vspace{-0.3cm}
\end{figure}

The photon number distribution $P(n)$ of the nonlinear coherent states is defined by \cite{Manko}
\begin{eqnarray}
P(n) = |\langle n| \alpha, \tilde{f}\rangle|^2.
\label{E1}
\end{eqnarray}
From (\ref{c13}) we find
\begin{eqnarray}
P(n) = \frac{\tilde{N}^2(|\alpha|^2)|\alpha|^{2n-6}}{\tilde{n}! [\tilde{f}(n)!]^2}.
\label{E2}
\end{eqnarray}
We plot the photon number distribution as a function of $n$ in Fig. \ref{first}(a).

To confirm the sub-Poissonian statistics exhibited by our non classical states,
we evaluate the Mandel parameter \cite{Mandel}
\begin{eqnarray}
Q = \frac{(\Delta\hat{n})^2 - \langle\hat{n}\rangle}{\langle \hat{n}\rangle} = \frac{\langle \hat{n}^2 \rangle - \langle\hat{n}\rangle^2-\langle \hat{n}\rangle}{\langle\hat{n}\rangle} = \frac{\langle \hat{n}^2 \rangle}{\langle\hat{n}\rangle}-\langle\hat{n}\rangle-1
\label{E3}
\end{eqnarray}
which should be negative $(Q < 0)$ \cite{Ant, Tava}.
First let us calculate $\langle \hat{n} \rangle$:
\begin{eqnarray}
\langle \hat{n} \rangle = \langle \alpha, \tilde{f}| \hat{n} |\alpha, \tilde{f} \rangle
                        = \tilde{N}^2 \sum^{\infty}_{n=3} \frac{12n|\alpha|^{2n-6}}{n!(n-1)!(n-3)!}.
\label{E4}
\end{eqnarray}
Substituting the value of $\tilde{N}$ (vide Eq. (\ref{c12})) in (\ref{E4}) and redefining $p = n-3$ in the
resultant expression, we get
\begin{eqnarray}
\langle \hat{n} \rangle =  \frac{\displaystyle\sum^{\infty}_{p=0} \frac{(p+3)|\alpha|^{2p}}{p!(p+2)!(p+3)!}}{\displaystyle\sum^{\infty}_{p=0}\frac{|\alpha|^{2p}}{p!(p+2)!(p+3)!}} = 3\left(\frac{{}_{1}F_3(4;3,3,4; |\alpha|^2) }{{}_{1}F_3(3;3,3,4; |\alpha|^2) }\right),
\label{E6}
\end{eqnarray}
where we have used the identity \cite{book}
\begin{eqnarray}
\fl {}_{p}F_q(\alpha_1,\alpha_2, ..., \alpha_p;\beta_1,\beta_1, \beta_1,...\beta_q;x) = \sum^{\infty}_{k=0} \frac{ (\alpha_1)_k (\alpha_2)_k...(\alpha_p)_k x^k}{ (\beta_1)_k (\beta_2)_k...(\beta_q)_k k!}, \quad (\alpha)_k = \frac{\Gamma(\alpha+k)}{\Gamma(\alpha)}.
\label{E7}
\end{eqnarray}
We plot the mean photon number $\langle \hat{n} \rangle$ as a function of $r(=|\alpha|^2)$
for the nonlinear coherent states (\ref{c13}) in Fig. \ref{first}(b).

In a similar fashion we find
\begin{eqnarray}
\langle \hat{n}^2\rangle = 9 \left(\frac{{}_{1}F_3(4;3,3,3; |\alpha|^2)}{{}_{1}F_3(3;3,3,4; |\alpha|^2)}\right).
\label{E10}
\end{eqnarray}
Substituting $\langle \hat{n}^2\rangle $ and $\langle \hat{n}\rangle $ in (\ref{E3}), we get
\begin{eqnarray}
Q = 3 \left(\frac{{}_{1}F_3(4;3,3,3; |\alpha|^2)}{{}_{1}F_3(4;3,3,4; |\alpha|^2)}-\frac{{}_{1}F_3(4;3,3,4; |\alpha|^2)}{{}_{1}F_3(3;3,3,4; |\alpha|^2)}\right)-1.
\label{E11}
\end{eqnarray}
The Mandel parameter $Q$ is depicted in Fig. \ref{first}(c) which confirms the sub-Poissonian statistics
established by the nonlinear coherent states for all values of $r$ $(= |\alpha|^2)$.

The quantity which determines bunching and anti-bunching of a state of the radiation field is simply decided by the second
order correlation function  \cite{Paul, Mah}, $g^{(2)}(0)$, which is defined by
\begin{eqnarray}
g^{(2)}(0) = \frac{\langle\hat{n}^2\rangle-\langle\hat{n}\rangle}{\langle \hat{n} \rangle^2}
           = \frac{\langle\hat{n}^2\rangle}{\langle \hat{n} \rangle^2} - \frac{1}{{\langle\hat{n}\rangle}}.
\label{E12}
\end{eqnarray}
Using (\ref{E6}) and (\ref{E10}) we find
\begin{eqnarray}
\fl \qquad g^{(2)}(0) = \left(\frac{{}_{1}F_3(3;3,3,4; |\alpha|^2)}{{}_{1}F_3(4;3,3,4; |\alpha|^2)}\right)\left(\frac{{}_{1}F_3(4;3,3,3; |\alpha|^2)}{{}_{1}F_3(4;3,3,4; |\alpha|^2)}-\frac{1}{3}\right).
\label{E13}
\end{eqnarray}

We depict the values of $g^{(2)}(0)$ in Fig. \ref{first}(d).
It is clear from Fig. \ref{first}(d) that since $g^{(2)}<1$ for $r>0$, the
Fock space states describing the light field is antibunched. In other words the nonlinear
coherent states form a Sub-Poissonian statistics.

Finally, we mention here that one can also express the mean photon number $\langle n \rangle$,
Mandel parameter $Q$ and the second order correlation function $g^{(2)}(0)$ in terms of
normalization constant $\tilde{N}(|\alpha|^2)$ as well \cite{Klauder2,Shree}. To implement this, we
redefine $|\alpha|^2 = y$ so that $\tilde{N}^2(|\alpha|^2) = {\cal N}(y)$. In this case
Eqs. (\ref{E6}), (\ref{E3}) and (\ref{E12}) can be expressed in terms of ${\cal N}(y)$ of the form
\begin{eqnarray}
\langle \hat{n} \rangle &=& y\frac{{\cal N}^{'}(y)}{{\cal N}(y)}+3, \\
Q &=& \left(\frac{y^{2}{\cal N}^{''}(y)+4y{\cal N}^{'}(y)}{y{\cal N}^{'}(y)+3{\cal N}(y)}\right) - y \frac{{\cal N}^{'}(y)}{{\cal N}(y)}-1, \\
g^{(2)}(0) &=& \left(\frac{{\cal N}(y)}{y{\cal N}^{'}(y)+3{\cal N}(y)}\right) \left(\frac{y^2{\cal N}^{''}(y)+4y{\cal N}^{'}(y)}{y{\cal N}^{'}(y)+3{\cal N}(y)} + 2\right).
\end{eqnarray}
respectively.

\section{Gazeau-Klauder coherent states}
\label{sec 5}
In this section, we construct Gazeau-Klauder coherent states \cite{Gazeau} for the Eq. (\ref{a2}).
To begin with, we consider the Hamiltonian $\tilde{H}$ satisfying the Schr\"{o}dinger equation
\begin{eqnarray}
\tilde{H}|n\rangle = e_n|n\rangle, \qquad e_n = nf^2(n).
\label{a42}
\end{eqnarray}
The Hamiltonian $\tilde{H}$ can be factorized in the number basis as
\begin{eqnarray}
\tilde{H}= \hat{B}^{\dagger}\hat{B},
\label{a43}
\end{eqnarray}
where $\hat{B}^{\dagger}$, $\hat{B}$ are the creation and annihilation operators, respectively defined by \cite{Kinani}
\begin{eqnarray}
\hat{B} = \hat{A} e^{\displaystyle i\gamma(\hat{e}_n-\hat{e}_{n-1})}, \;
\hat{B}^{\dagger} = e^{\displaystyle -i\gamma(\hat{e}_{n}-\hat{e}_{n-1})}\hat{A}^{\dagger}, \quad  \gamma \in \mathbb{R}.
\label{a44a}
\end{eqnarray}
They act on the states $|n\rangle$ as
\begin{eqnarray}
\hat{B}|n\rangle = \sqrt{e_n} e^{i\gamma(e_n-e_{n-1})} |n-1\rangle, \;\;\;
\hat{B}^{\dagger}|n\rangle = \sqrt{e_{n+1}} e^{-i\gamma(e_{n+1}-e_n)}| n+1\rangle.
\label{a44}
\end{eqnarray}

The Gazeau-Klauder coherent states $|z, \gamma\rangle$
are defined as the eigenstates of annihilation operator $\hat{B}$. They
satisfy the eigenvalue equation \cite{Gazeau}
\begin{eqnarray}
\hat{B}|z, \gamma\rangle = z|z, \gamma\rangle,  \qquad z \in \mathbb{C}.
\label{a45}
\end{eqnarray}
Substituting
\begin{eqnarray}
|z, \gamma\rangle = \sum^{\infty}_{n=0} c_n |n\rangle, \qquad n \neq 1,2,
\label{a46}
\end{eqnarray}
in Eq. (\ref{a45}) and equating the coefficients of $\psi_n$ in the resultant equation we get
\begin{eqnarray}
c_{n+1} \sqrt{e_{n+1}} e^{i\gamma(e_{n+1}-e_n)} = z c_n.
\label{a47}
\end{eqnarray}
From (\ref{a47}), we find $ c_0 = 0 $ and one can express $c_n$ in terms of $c_3$ as
\begin{eqnarray}
c_n = \frac{z^{n-3}e^{-i\gamma e_n}}{\sqrt{\tilde{n}!}\;\tilde{f}(n)!}c_3,
\label{a48}
\end{eqnarray}
with $e_{n} = n(n-1)(n-3)$.
We mention here that the coherent states defined for the Hamiltonian (\ref{a42}) do not contain the states with photon number
less than three. The coherent states contain the states with photon number starting
from three reads now
\begin{eqnarray}
\tilde{|z, \gamma\rangle} = \tilde{N}(|z|^2) \sum^{\infty}_{n=3}\frac{z^{n-3}e^{-i\gamma e_n}}{\sqrt{\tilde{n}!}\;\tilde{f}(n)!}|n\rangle.
\label{a49}
\end{eqnarray}
To determine $\tilde{N}(|z|^2)$, we use the normalization condition, $\langle \tilde{z, \gamma}|\tilde{z, \gamma}\rangle = 1$,
which in turn yields
\begin{eqnarray}
\tilde{N}(|z|^2) = \left(\sum^{\infty}_{n=3}\frac{|z|^{2n-6}}{\tilde{n}!\;[\tilde{f}(n)!]^2}\right)^{-1/2}.
\label{a51}
\end{eqnarray}

It is clear that the nonlinear coherent states are continuous in $z\in \mathbb{C}$ and are temperorally
stable under the evolution operator.  The latter can be confirmed by verifying the condition
\begin{eqnarray}
e^{-i\tilde{H}t}|\tilde{z,\gamma}\rangle = |\tilde{z, \gamma+t}\rangle.
\label{a52}
\end{eqnarray}

To confirm the Gazeau-Klauder coherent states resolve the identity one must find a measure $d\mu(z) = W(|z|^{2}) d^{2}z$ such that \cite{Klauder2}
\begin{eqnarray}
\frac{1}{\pi} \int \int_\mathbb{C} |\tilde{z,\gamma}\rangle W(|z|^2)\langle \tilde{z,\gamma}| d^{2}z = \hat{I},
\label{a53}
\end{eqnarray}
where the integration should be carried over the entire complex plane.
Now let us evaluate $\langle\psi|\hat{I}|\Phi\rangle$. From (\ref{a53}) we have
\begin{eqnarray}
\frac{1}{\pi} \int \int_\mathbb{C} \langle\psi\tilde{|z, \gamma\rangle} W(|z|^2)\tilde{\langle z, \gamma|}\Phi\rangle d^2 z = \langle\psi|\hat{I}|\Phi\rangle.
\label{aa55}
\end{eqnarray}
Substituting $\tilde{|z,\gamma\rangle}$ and its conjugate in (\ref{aa55}), we get
\begin{eqnarray}
\fl \qquad \frac{1}{\pi} \sum^{\infty}_{m,n=3}\frac{\langle\psi|n\rangle \langle m|\Phi\rangle}{\sqrt{\tilde{n}!\tilde{m}!}\; \tilde{f}(n)!\tilde{f}(m)!}  \int \int_\mathbb{C} z^{n-3}{z^{*}}^{m-3} \tilde{N}^{2}(|z|^2) W(|z|^2)d^2 z  = \langle\psi|\hat{I}|\Phi\rangle.
\label{aa56}
\end{eqnarray}

The integral on the left hand side in Eq. (\ref{aa56}) can be evaluated along
the same lines as given in Eqs. (\ref{d5}) - (\ref{d7}). As a result one
can bring the left hand side of Eq. (\ref{aa56}) to the form
$\sum^{\infty}_{n=3}\langle\psi|n\rangle\langle n|\Phi \rangle$. In this case also, the final
expression turns out to be a projector for which the the ground state $|0\rangle$ is excluded.

Using Eq. (\ref{a45}), one can obtain the action identity \cite{Gazeau}
\begin{eqnarray}
\langle \tilde{z, \gamma}|\tilde{H}|\tilde{z, \gamma}\rangle = |z|^{2}.
\label{a55}
\end{eqnarray}
The function $e_n=nf^2(n)$ has zeroes at $n = 0, 1$ and $3$. Since we have zeroes in the operators,
the Fock space breaks up into a countable number of irreducible representations.
As cited in  \cite{Klauder} if the zeroes are simple zeroes, some of the reduced pieces do not allow a unitary representation
and the associated coherent states cannot be expressed as Klauder-Perelomov's one.

We find that the photon number distribution $P(n)$ for the Gazeau-Klauder coherent states of
the nonlinear oscillator (\ref{a2}) in the form
\begin{eqnarray}
P(n) = |\langle n\tilde{| z, \gamma}\rangle|^2 = \frac{\tilde{N}^2(|z|^2)|z|^{2n-6}}{\tilde{n}! [\tilde{f}(n)!]^2},
\label{P1}
\end{eqnarray}
where we have used (\ref{a49}) to calculate $P(n)$. The Mandel parameter, $Q$,
and the second  order correlation function, $g^{(2)}(0)$, turns out to be
\begin{eqnarray}
  Q &=& 3 \left(\frac{{}_{1}F_3(4;3,3,3; |z|^2)}{{}_{1}F_3(4;3,3,4; |z|^2)}-\frac{{}_{1}F_3(4;3,3,4; |z|^2)}{{}_{1}F_3(3;3,3,4; |z|^2)}\right)-1,  \\
  g^{(2)}(0) &=& \left(\frac{{}_{1}F_3(3;3,3,4; |z|^2)}{{}_{1}F_3(4;3,3,4; |z|^2)}\right)\left(\frac{{}_{1}F_3(4;3,3,3; |z|^2)}{{}_{1}F_3(4;3,3,4; |z|^2)}-\frac{1}{3}\right)
\label{p11}
\end{eqnarray}
respectively. For $|z|>0 $ we find $Q<0$ and these negative values of $Q$ indicate that
the Gazeau-Klauder coherent states possess sub-Poissonian distribution.
Again for $|z|>0$ we find $g^{(2)}(0) < 1$ which in turn confirms
the anti-bunching of the light field.

\section{Even and odd nonlinear coherent states}
\label{sec6}
In this section, we construct even and odd nonlinear coherent states \cite{Man} for the Schr\"{o}dinger equation (\ref{a2}).
The even and odd nonlinear coherent states are the symmetric and antisymmetric combination of the nonlinear
coherent states. They are two orthonormalized eigenstates of the square of the
annihilation operator and essentially have two kinds of nonclassical effects: the even nonlinear coherent states
have a squeezing but no anti-bunching while the odd nonlinear coherent states
have an anti-bunching but no squeezing.
The even and odd nonlinear coherent states \cite{Siva} defined as the eigenstates of $\hat{A}^2$:
\begin{eqnarray}
\hat{A}^2|\psi\rangle = \alpha |\psi\rangle.
\label{a56}
\end{eqnarray}

Substituting $ |\psi\rangle = \sum^{\infty}_{n=0}c_n|n\rangle, \;n \neq 1,2$ in Eq. (\ref{a56}) and expanding we get
\begin{eqnarray}
\sum^{\infty}_{n=0}c_n\sqrt{n(n-1)}f(n)f(n-1)|n-2\rangle = \alpha \sum^{\infty}_{n=0}c_n|n\rangle.
\label{a58}
\end{eqnarray}
From (\ref{a58}) we find the constant $c_0 = 0$.

To determine the value of the rest of the coefficients we proceed as follows.
Equating the coefficients of $\psi_{n}$ in Eq. (\ref{a58})
and denoting
\begin{eqnarray}
F(n) = \sqrt{n+1}f(n+1)f(n+2)
\label{a58a}
\end{eqnarray}
in that expression we find a relation between $c_{n+2}$ and $c_n$ of the form
\begin{eqnarray}
\sqrt{n+2}\;F(n)c_{n+2} = \alpha c_n,
\label{a59}
\end{eqnarray}
from which we obtain
\begin{eqnarray}
\sqrt{n}\;F(n-2)c_{n} = \alpha c_{n-2}.
\label{a59a}
\end{eqnarray}
In the case $n$ is an even number, we can redefine $n = 2n$ and fix
\begin{eqnarray}
c_{2n} = \frac{\alpha}{\sqrt{2n}\; F(2n-2)}c_{2n-2}.
\label{a60}
\end{eqnarray}
Evaluating $c_{2n-2}$ in terms of $c_{2n-4}$ and so on, one can find the value
of the even number coefficients as
\begin{eqnarray}
\fl \qquad c_{2n} = \frac{\alpha^{n-2}}{\sqrt{2n.(2n-2).(2n-4)...6}\; F(2n-2)F(2n-4)F(2n-6)...F(4)}c_4.
\label{a61}
\end{eqnarray}

On the other hand $n$ is an odd number we can redefine $n=2n+1$ and fix
\begin{eqnarray}
c_{2n+1} = \frac{\alpha}{\sqrt{2n+1} F(2n-1)}c_{2n-1}.
\label{a62}
\end{eqnarray}
Now evaluating $c_{2n-1}$ in terms of $c_{2n-3}$ and so on we
find the value of the odd number coefficients as
\begin{eqnarray}
\fl \;\;\;\; \qquad c_{2n+1} = \frac{\alpha^{n-2}}{\sqrt{(2n+1)(2n-1)(2n-3)...5}\;F(2n-1)F(2n-3)...F(3)}c_3.
\label{a63}
\end{eqnarray}
For simplicity we redefine the constants $c_{2n}$ and $c_{2n+1}$ as
\begin{eqnarray}
\fl \qquad c_{2n} = \frac{\alpha^{n-2}}{\sqrt{\widetilde{(2n)}!!}\;\tilde{F}(2n-2)!!}c_4, \quad
c_{2n+1}= \frac{\alpha^{n-2}}{\sqrt{(\widetilde{2n+1})!!}\;\tilde{F}(2n-1)!!}c_3,
\label{a64}
\end{eqnarray}
where
\begin{eqnarray}
\widetilde{(2n)}!! &=& 2n. (2n-2). (2n-4)...6, \\
\tilde{F}(2n-2)!! &=& F(2n-2)F(2n-4)F(2n-6)...F(4),\\
\widetilde{(2n+1)}!! &=& (2n+1). (2n-1). (2n-3)...5,\\
\tilde{F}(2n-1)!! &=& F(2n-1)F(2n-3)...F(3).
\end{eqnarray}

As a result, one gets even and odd nonlinear coherent states for the Eq. (\ref{a2}) respectively of the form
\begin{eqnarray}
|\alpha, \tilde{F}, +\rangle &=& c_4\sum^{\infty}_{n=2}\frac{\alpha^{n-2}}{\sqrt{\widetilde{(2n)}!!}\;\tilde{F}(2n-2)!!}\;|2n\rangle,
\label{a66}
\end{eqnarray}
\begin{eqnarray}
|\alpha, \tilde{F}, -\rangle &=& c_3\sum^{\infty}_{n=1}\frac{\alpha^{n-1}}{\sqrt{\widetilde{(2n+1)!!}}\;\tilde{F}(2n-1))!!}\;|2n+1\rangle,
\label{a66a}
\end{eqnarray}
where $\tilde{F}(1)!! = \tilde{3}!! = 1$.

The constants $c_4 (= N_e)$ and $c_3 (= N_o)$ can be fixed easily through the normalization procedure
$\langle \alpha, \tilde{F}, \pm |\alpha, \tilde{F}, \pm \rangle = 1, $
which turns out to be
\begin{eqnarray}
\fl N_e =  \left(\sum^{\infty}_{n=2}\frac{|\alpha|^{2n-4}}{\widetilde{(2n)}!!\;[\tilde{F}(2n-2)!!]^2}\right)^{-1/2},
N_o = \left(\sum^{\infty}_{n=1}\frac{|\alpha|^{2n-2}}{\widetilde{(2n+1)}!!\;[\tilde{F}(2n-1))!!]^2}\right)^{-1/2}.
\label{a68}
\end{eqnarray}

The photon number distribution  and mean photon number of even nonlinear coherent states, turns out to be
\begin{eqnarray}
\langle \hat{n} \rangle = N = 4\left(\frac{{}_{0}F_5(;2,2,\frac{3}{2},\frac{5}{2},\frac{5}{2}; \frac{|\alpha|^2}{64})}{{}_{0}F_5(;2,3,\frac{3}{2},\frac{5}{2},\frac{5}{2};  \frac{|\alpha|^2}{64})}\right) \nonumber\\
P(2n) = |\langle 2n|\alpha, \tilde{F}, +\rangle|^2 = \frac{N^{2}_e|\alpha|^{2n-4}}{\tilde{(2n)}!!\;[\tilde{F}(2n-2)!!]^2}.
\label{a69}
\end{eqnarray}

We find the Mandel parameter, $Q$ and the second  order correlation function, $g^{(2)}(0)$ of even nonlinear coherent states of the form
\begin{eqnarray}
\fl \qquad Q & = 4\left(\frac{{}_{1}F_6(3;2,2,2, \frac{3}{2},\frac{5}{2},\frac{5}{2}; \frac{|\alpha|^2}{64})}{{}_{0}F_5(;2,2,\frac{3}{2},\frac{5}{2},\frac{5}{2};  \frac{|\alpha|^2}{64})}\right) - 4\left(\frac{{}_{0}F_5(;2,2,\frac{3}{2},\frac{5}{2},\frac{5}{2}; \frac{|\alpha|^2}{64})}{{}_{0}F_5(;2,3,\frac{3}{2},\frac{5}{2},\frac{5}{2};  \frac{|\alpha|^2}{64})}\right) - 1  ,\nonumber \\
\fl \qquad g^{2}(0) & = \left(\frac{{}_{0}F_5(;2,3,\frac{3}{2},\frac{5}{2},\frac{5}{2}; \frac{|\alpha|^2}{64})}{{}_{0}F_5(;2,2,\frac{3}{2},\frac{5}{2},\frac{5}{2};  \frac{|\alpha|^2}{64})}\right)\left(\frac{{}_{1}F_6(3;2,2,2,\frac{3}{2},\frac{5}{2},\frac{5}{2}; \frac{|\alpha|^2}{64})}{{}_{0}F_5(;2,2,\frac{3}{2},\frac{5}{2},\frac{5}{2};  \frac{|\alpha|^2}{64})}- \frac{1}{4}\right)
\label{a69a}
\end{eqnarray}
respectively.

\begin{figure}[!ht]
\vspace{-0.4cm}
\begin{center}
\includegraphics[width=0.95\linewidth]{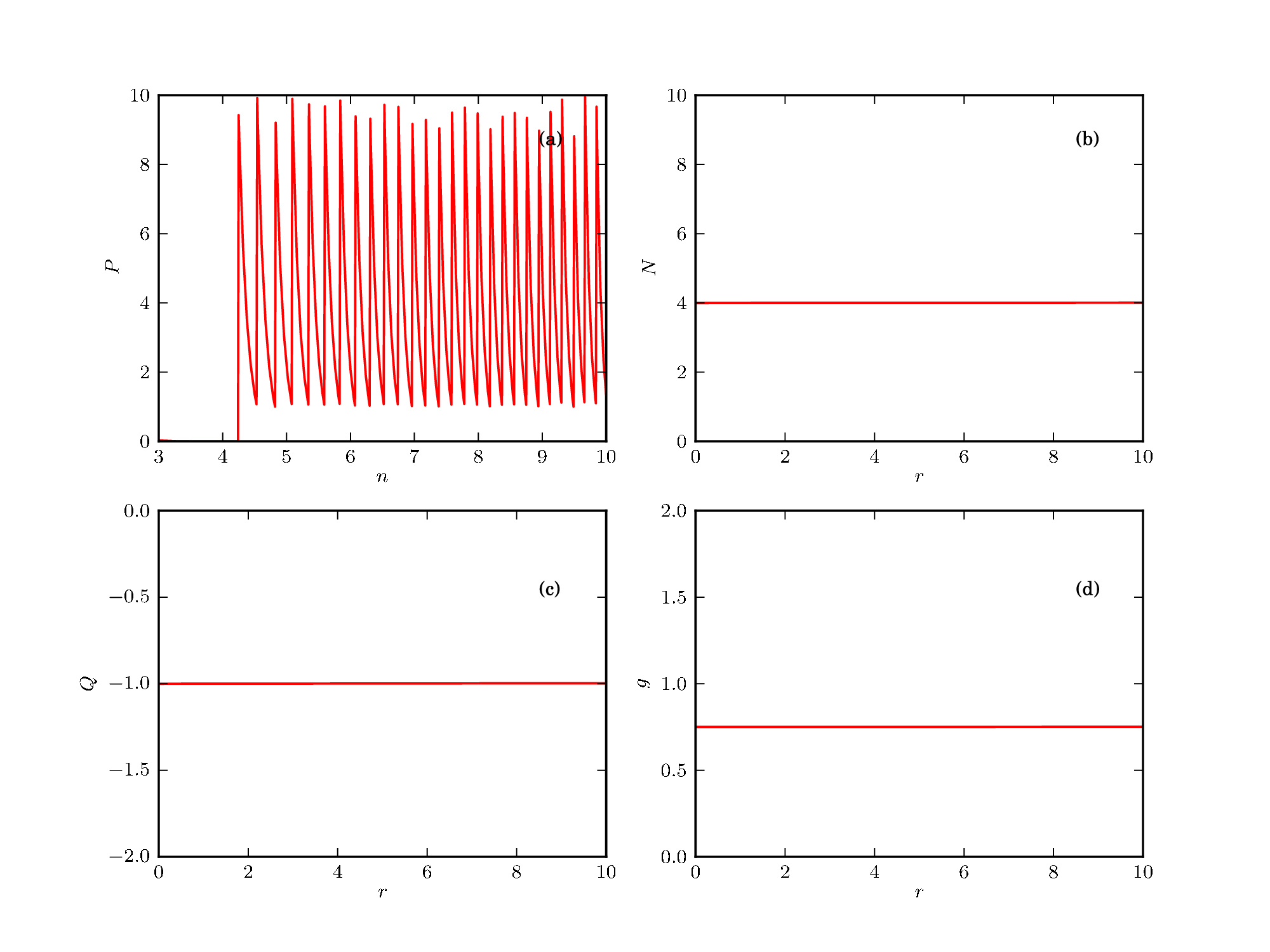}
\end{center}
\vspace{-1cm}
\caption{The plots of (a) photon number distribution $P(n)$, (b) mean photon number $(\langle \hat{n} \rangle)$, (c) Mandel parameter $Q$
and (d) the second order correlation function $g^{2}(0)$ (denoted as $g$) of even nonlinear coherent states (\ref{a66}).} \label{second}
\vspace{-0.3cm}
\end{figure}

We plot the mean photon number $\langle \hat{n} \rangle$, Mandel parameter, $Q$, and
the second  order correlation function, $g^{(2)}(0)$
as a function of $r (=(|\alpha|^2)$ for the  even nonlinear coherent states
(\ref{a66}) in Fig. \ref{second}(a),(b),(c) and (d) respectively.
The Mandel parameter $Q$ and the second  order correlation function, $g^{(2)}(0)$
depicted in Fig. \ref{second}(c) and (d) confirm the sub-Poissonian statistics
established by the even nonlinear coherent states for all positive values of $r$ $(= |\alpha|^2)$.

\begin{figure}[!ht]
\vspace{-0.4cm}
\begin{center}
\includegraphics[width=0.95\linewidth]{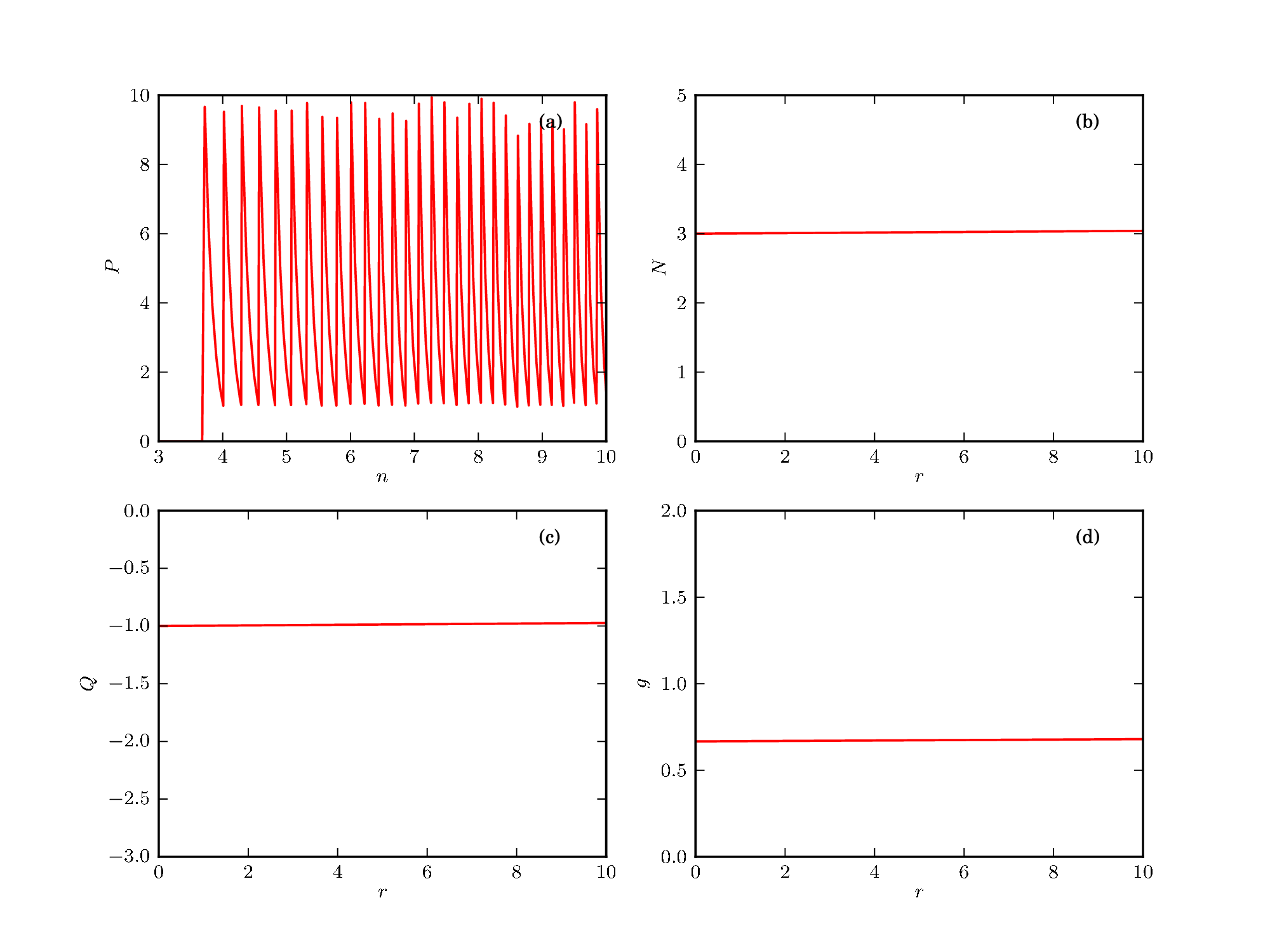}
\end{center}
\vspace{-1cm}
\caption{The plots of (a) photon number distribution $P(n)$ , (b) mean photon number $(\langle \hat{n} \rangle)$, (c) Mandel parameter $Q$ and (d) the second order correlation function $g^{2}(0)$ (denoted as $g$) of odd nonlinear coherent states (\ref{a66a}).} \label{third}
\vspace{-0.3cm}
\end{figure}

Whereas photon number distribution, mean photon number, Mandel parameter, $Q$, and
the second  order correlation function, $g^{(2)}(0)$ of odd nonlinear coherent states, turns out to be
\begin{eqnarray}
\langle \hat{n} \rangle = N = 3\left(\frac{{}_{0}F_5(;2,2,\frac{1}{2},\frac{3}{2},\frac{3}{2}; \frac{|\alpha|^2}{64})}{{}_{0}F_5(;2,2,\frac{1}{2},\frac{3}{2},\frac{5}{2};  \frac{|\alpha|^2}{64})}\right), \nonumber \\
P(2n+1) = |\langle 2n+1|\alpha, \tilde{F}, -\rangle|^2 = \frac{N^2_o|\alpha|^{2n-2}}{\tilde{(2n+1)}!!\;[\tilde{F}(2n-1)!!]^2},\nonumber\\
 Q = 3\left(\frac{{}_{1}F_6(\frac{5}{2};2,2,\frac{1}{2},\frac{3}{2},\frac{3}{2},\frac{3}{2}; \frac{|\alpha|^2}{64})}{{}_{0}F_5(;2,2,\frac{1}{2},\frac{3}{2},\frac{3}{2};  \frac{|\alpha|^2}{64})}\right) - 3\left(\frac{{}_{0}F_5(;2,2,\frac{1}{2},\frac{3}{2},\frac{3}{2}; \frac{|\alpha|^2}{64})}{{}_{0}F_5(;2,2,\frac{1}{2},\frac{3}{2},\frac{5}{2};  \frac{|\alpha|^2}{64})}\right) - 1  ,\nonumber \\
  g^{2}(0) = \left(\frac{{}_{0}F_5(;2,2,\frac{1}{2},\frac{3}{2},\frac{5}{2}; \frac{|\alpha|^2}{64})}{{}_{0}F_5(;2,2,\frac{1}{2},\frac{3}{2},\frac{3}{2};  \frac{|\alpha|^2}{64})}\right)\left(\frac{{}_{1}F_6(\frac{5}{2};2,2,\frac{1}{2},\frac{3}{2},\frac{3}{2},\frac{3}{2}; \frac{|\alpha|^2}{64})}{{}_{0}F_5(;2,2,\frac{1}{2},\frac{3}{2},\frac{3}{2};  \frac{|\alpha|^2}{64})}- \frac{1}{3}\right),
\label{a71}
\end{eqnarray}
respectively.
We plot the mean photon number $\langle \hat{n} \rangle$, Mandel parameter, $Q$, and
the second  order correlation function, $g^{(2)}(0)$
as a function of $r (=|\alpha|^2)$ for the  odd nonlinear coherent states
(\ref{a71}) in Fig. \ref{third}(a),(b),(c) and (d) respectively.
The Mandel parameter $Q$ and the second  order correlation function, $g^{(2)}(0)$
depicted in Fig. \ref{third}(c) and (d) confirm the sub-Poissonian statistics
established by the odd nonlinear coherent states for the values of $r > 0$.

\section{$f$-deformed coherent states for position dependent mass nonlinear oscillator}
\label{sec8}

Using point canonical transformation method, Kraenkel and Senthilvelan \cite{Sen},
have shown that the position dependent mass Schr\"{o}dinger equation
\begin{eqnarray}
-\frac{1}{2}\frac{d}{dy}\left(\frac{1}{m(y)}\frac{d\tilde{\psi}_n(y)}{dy}\right)+\tilde{V}(y)\tilde{\psi}_n(y) = \tilde{E}_n\tilde{\psi}_n(y),
\label{n1}
\end{eqnarray}
where
\begin{eqnarray}
\fl \qquad \tilde{V}(y) = \frac{1}{2}\left(\eta^2+\frac{8(2\eta^2 - 1)}{(1+2\eta^2)^2} + 3\right)+\frac{m^{''}}{8m^2}-\frac{7{m^{'}}^{2}}{32m^{3}},
\quad \frac{d\eta}{dy} \le m^{1/2}(y)
\label{n2}
\end{eqnarray}
is also exactly solvable. In the above $m(y)$ is the mass distribution function which depends on the coordinate $y$.
The authors have obtained the eigenfunctions and energy eigenvalues of (\ref{n1}) in the form
\begin{eqnarray}
\fl \qquad \tilde{\psi}_n(y) &= \tilde{{\cal N}}_n m^{1/4}(y)\frac{{\cal P}_n(\eta(y))}{(1+2y^2)} e^{\displaystyle -\frac{\eta^2}{2}},\;
 E_{n} &=  \tilde{E}_n = -\frac{3}{2}+n,\; n=0,3,4,\ldots,
\label{n3}
\end{eqnarray}

For the sake of illustration the authors have considered three different types of
mass distributions which are often being considered in semiconductor physics and
constructed bound state energies and wavefunctions for these cases.

In the following, we consider the position dependent
mass Schr\"{o}dinger equation (\ref{n1}) associated with the nonlinear oscillator and construct its nonlinear
coherent states.

To determine the deformed ladder operators ${\cal \hat{A}}$, ${\cal \hat{A}}^{\dagger}$
we recall the definitions
\begin{eqnarray}
{\cal \hat{A}}\tilde{\psi}_n &= c_1(n)\tilde{\psi}_{n-1},
\label{n6}
\end{eqnarray}
\begin{eqnarray}
{\cal \hat{A}}^{\dagger}\tilde{\psi}_n &= c_2(n)\tilde{\psi}_{n+1},
\label{n7}
\end{eqnarray}
with $\tilde{\psi}_n$ given in (\ref{n3}). Assuming, ${\cal \hat{A}}$ and ${\cal \hat{A}}^{\dagger}$
are of the same forms (\ref{a13}) and (\ref{a17}) and demanding that they should
satisfy the conditions (\ref{n6}) and (\ref{n7}) we find that the
deformed ladder operators ${\cal \hat{A}}$ and ${\cal \hat{A}}^{\dagger}$
be of the form
\begin{eqnarray}
\fl \qquad \sqrt{2}{\cal \hat{A}} &=& \left[\frac{2(2\eta^2 - 1)}{(1+2\eta^2)^2}-1\right]\left[\frac{1}{\sqrt{m(y)}}\frac{d}{dy} + \phi\right]+\left[\frac{1}{\sqrt{m(y)}}\frac{d}{dy} + \eta \right]\hat{n},\\
\fl \qquad \sqrt{2}{\cal \hat{A}}^{\dagger} &=& \left[\frac{2(1-2\eta^2)}{(1+2\eta^2)^2}\right]\left[\frac{1}{\sqrt{m(y)}}\frac{d}{dy} + \phi\right]+\left[-\frac{1}{\sqrt{m(y)}}\frac{d}{dy} + \eta \right]\hat{n}.
\label{n8}
\end{eqnarray}

We mention here that to determine these two operators one needs to derive
two recurrence relations in the ${\cal P}$-Hermite polynomials of the form
\begin{eqnarray}
\fl \qquad \left[n-1+\frac{2(2\eta^2 - 1)}{(1+2\eta^2)^2}\right]{\cal P}_n^{'}(\eta(y))+n(\eta-\phi){\cal P}_n(\eta(y)) = 2n(n-3){\cal P}_{n-1}(\eta(y)),
\label{n4}
\end{eqnarray}
\begin{eqnarray}
\fl \quad-\left[n+\frac{2(2\eta^2-1)}{(1+2\eta^2)^2}\right]{\cal P}^{'}_n(\eta(y)) + n(\eta+\phi){\cal P}_n(\eta(y)) = n{\cal P}_{n+1}(\eta(y)),
\label{n5}
\end{eqnarray}
where  $\phi = \eta + \frac{4\eta}{(1+2\eta^2)}$ and $\eta(y)$ is given in Eq. (\ref{n2}).

It is now a simple matter to verify
\begin{eqnarray}
{\cal \hat{A}} \tilde{\psi}_{n}&= \sqrt{n}\;f(n)\;\tilde{\psi}_{n-1}, \\
{\cal \hat{A}}^{\dagger} \tilde{\psi}_{n} &= \sqrt{n+1}\;f(n+1)\; \tilde{\psi}_{n+1},
\label{n9}
\end{eqnarray}
with $f(n) = \sqrt{(n-1)(n-3)}$.

Since the deformed ladder operators for the PDMSE (\ref{n1}) are also of the same form as in the case of constant mass Schr\"{o}dinger
equation, except the presence of position dependent mass terms, the intelligent states and nonlinear coherent states for Eq. (\ref{n1})
can be constructed in the same way as in the constant mass case.
The resultant expressions read
\begin{eqnarray}
|\alpha, \tilde{f}, \lambda\rangle = c_3\sum^{\infty}_{n=3}\frac{\tilde{A}_n !}{\sqrt{\tilde{n}!}\tilde{f}(n)!}\tilde{\psi}_{n}.
\label{n10}
\end{eqnarray}
and
\begin{eqnarray}
|\alpha, \tilde{f}\rangle = \tilde{N} \sum^{\infty}_{n=3}\frac{\alpha^{n-3}}{\sqrt{\tilde{n}!}\tilde{f}(n)!}\tilde{\psi}_{n}.
\label{n10a}
\end{eqnarray}
In the above, the arbitrary constant $\tilde{N}$, can be fixed by the normalization procedure
which turns out to be
\begin{eqnarray}
\tilde{N} = \left(\sum^{\infty}_{n=3}\frac{|\alpha|^{2n-6}}{\tilde{n}!\; [\tilde{f}(n)!]^2}\right)^{-1/2}.
\label{n11a}
\end{eqnarray}

The Gazeau-Klauder coherent states for position dependent mass Schr\"{o}dinger
equation (\ref{n2}) read
\begin{eqnarray}
\tilde{|z, \gamma\rangle} = \tilde{N}(|z|^2) \sum^{\infty}_{n=3}\frac{z^{n-3}}{\sqrt{\tilde{n}!}\;\tilde{f}(n)!}\tilde{\psi}_{n}.
\label{n11}
\end{eqnarray}
with
\begin{eqnarray}
\tilde{N}(|z|^2) = \left(\sum^{\infty}_{n=3}\frac{|z|^{2n-6}}{\tilde{n}!\;[\tilde{f}(n)!]^2}\right)^{-1/2}.
\label{n12}
\end{eqnarray}

The even and odd nonlinear coherent states for the Eq. (\ref{n2}) can also be constructed along the same line
as given in the constant mass case. The resultant expressions read
\begin{eqnarray}
|\alpha, \tilde{F}, +\rangle &=& N_e\sum^{\infty}_{n=2}\frac{\alpha^{n-2}}{\sqrt{\widetilde{(2n)}!!}\;\widetilde{F}(2n-2)!!}\;
\tilde{\psi}_{2n},
\label{n12a}
\end{eqnarray}
\begin{eqnarray}
|\alpha, \tilde{F}, -\rangle &=& N_o\sum^{\infty}_{n=1}\frac{\alpha^{n-1}}{\sqrt{\widetilde{(2n+1)!!}}\;\widetilde{F}(2n-1))!!}\;
\tilde{\psi}_{2n+1},
\label{n13}
\end{eqnarray}
where the normalization constants $N_e$ and $N_e$ are given by
\begin{eqnarray}
\fl N_e = \left(\sum^{\infty}_{n=2}\frac{|\alpha|^{2n-4}}{\widetilde{(2n)}!!\;[\tilde{F}(2n-2)!!]^2}\right)^{-1/2},
N_o = \left(\sum^{\infty}_{n=1}\frac{|\alpha|^{2n-2}}{\widetilde{(2n+1)}!!\;[\tilde{F}(2n-1))!!]^2}\right)^{-1/2},
\label{n14}
\end{eqnarray}
respectively.

\subsection{Example:}

In the following,  we consider a specific mass profile and
give the explicit forms of the nonlinear coherent states for the
position dependent mass Schr\"{o}dinger equation (\ref{n2}).
Let us  consider the mass profile \cite{pdms}
\begin{eqnarray}
m(y) = \frac{(\gamma + y^2)^2}{(1+y^2)^2}, \qquad \gamma = constant,
\label{n15}
\end{eqnarray}
which is found to be useful for studying transport
properties in semiconductors \cite{Koc,Miller}. We also note that
the normalization constant is the same as the constant mass Schr\"{o}dinger equation (\ref{a2}).
Substituting (\ref{n15}) in the Eq. (\ref{n1})
we get
\begin{eqnarray}
\fl \left[-\frac {1}{2}\frac {d}{dy}\left(\frac{(1+y^2)^2}{(\gamma + y^2)^2}\frac{d}{dy}\right)+ \frac{1}{2}\left(\eta^2+\frac{8(2\eta^2 - 1)}{(1+2\eta^2)^2}+3 \right)+\frac{(\gamma-1)(3y^4+2(2-\gamma)y^2-\gamma)}{2(\gamma+y^2)^4}\right]\tilde{\psi}_n(y)\, \nonumber \\
 = \tilde{E}_n\tilde{\psi}_n(y),
\label{n16}
\end{eqnarray}
where $\eta(y)$ is given by
\begin{eqnarray}
\eta(y) = \int m^{\frac{1}{2}}(y) dy = y+(\gamma-1)tan^{-1}y,
\quad -\infty < \eta(y)<\infty .
\label{n17}
\end{eqnarray}
The coherent states given in Eqs. (\ref{n10}), (\ref{n10a}), (\ref{n11}), (\ref{n12a}) and (\ref{n13})
yield
\begin{eqnarray}
\fl \quad |\alpha, f, \lambda\rangle &= &c_3 \sum^{\infty}_{n=3}\frac{\tilde{A}_n !}{\sqrt{\tilde{n}!}\tilde{f}(n)!}\left(\frac{(\gamma + y^2)^2}{(1+y^2)^2}\right)^{1/4} \frac{{\cal P}_n(\eta)e^{-\frac{\eta^2}{2}}}{(1+2\eta^2)}, \\
\fl \quad|\alpha, \tilde{f}\rangle & = & \tilde{N} \sum^{\infty}_{n=3}\frac{\alpha^{n-3}}{\sqrt{\tilde{n}!}\tilde{f}(n)!}
\left(\frac{(\gamma + y^2)^2}{(1+y^2)^2}\right)^{1/4} \frac{{\cal P}_n(\eta)e^{-\frac{\eta^2}{2}}}{(1+2\eta^2)}, \\
\fl \quad\tilde{|z, \gamma\rangle} & = & \tilde{N}(|z|^2) \sum^{\infty}_{n=3}\frac{z^{n-3}e^{-i\gamma e_n}}{\sqrt{\tilde{n}!}\;\tilde{f}(n)!}
\left(\frac{(\gamma + y^2)^2}{(1+y^2)^2}\right)^{1/4}\frac{{\cal P}_n(\eta)e^{-\frac{\eta^2}{2}}}{(1+2\eta^2)}, \\
\fl \quad|\alpha, \tilde{F}, +\rangle &=& N_e\sum^{\infty}_{n=2}\frac{\alpha^{n-2}}{\sqrt{\widetilde{(2n)}!!}\;\widetilde{F}(2(n-1))!!}\;
\left(\frac{(\gamma + y^2)^2}{(1+y^2)^2}\right)^{1/4}\frac{{\cal P}_{2n}(\eta)e^{-\frac{\eta^2}{2}}}{(1+2\eta^2)}, \\
\fl \quad |\alpha, \tilde{F}, -\rangle &=& N_o\sum^{\infty}_{n=1}\frac{\alpha^{n-1}}{\sqrt{\widetilde{(2n+1)!!}}\;\widetilde{F}(2n-1))!!}\;
\left(\frac{(\gamma + y^2)^2}{(1+y^2)^2}\right)^{1/4} \frac{{\cal P}_{2n+1}(\eta)e^{-\frac{\eta^2}{2}}}{(1+2 \eta^2)}.
\label{n10b}
\end{eqnarray}
where $\tilde{N}(|\alpha|^2)$, $\tilde{N}(|z|^2)$, $N_e$ and $N_o$  are given in
Eqs. (\ref{n11a}), (\ref{n12}) and (\ref{n14}) respectively.

\section{Conclusion}
In this paper, we have considered the newly solvable nonlinear oscillator
that is related to the isotonic oscillator and constructed nonlinear coherent states for it.
Deviating from the conventional way, the ladder operators for this exactly solvable quantum
system have been derived from the solution of the Schr\"{o}dinger equation.
Two recursion relations involving ${\cal P}$-Hermite polynomial
have been derived to obtain the ladder operators. The same methodology has been adopted
to explore the generalized/deformed annihilation and
creation operators for the position dependent mass Schr\"{o}dinger equation also.
We have also shown that these operators satisfy the Heisenberg
algebra in the Fock space. We found that the nonlinearity function $f(\hat{n})$ in the Fock space
becomes zero at two values. As a consequence, the generalized intelligent states and nonlinear coherent states
constructed by us decomposes into two sub-sets in the Fock space. We have also constructed
the generalized intelligent states, Gazeau-Klauder coherent states
and even and odd nonlinear coherent states for this system.
Further, we have studied the Hilbert space properties and certain nonclassical properties of
these nonlinear coherent states. In addition to the above, we have considered position dependent
mass Schr\"{o}dinger equation associated with the Eq. (\ref{a2}) and
constructed nonlinear coherent states, even and odd nonlinear coherent states, Gazeau-Klauder
coherent states for it. This is motivated from the fact that in a wide
variety of physical problems an effective mass depending on the position is of utmost relevance.
To cite one such situation, in the nanofabrication of semiconductor devices, one observes quantum wells with
very thin layers. The effective mass of an electron hole in the thin layered quantum wells varies
with the composition rate. In such systems, the mass of the electron may change with the composition
rate, which depends on the position. As a consequence, attempts have been made to
analyze such PDMSE and their underlying properties for a number of potentials and masses. One
such mass profile which is found to be useful for studying transport properties in 
semiconductors is (\ref{n15}).
We have given explicit expressions for these nonlinear coherent states
for this mass profile.
The results obtained in this paper will give deep insights into geometrical
properties of the nonlinear oscillator potential (\ref{a1}).

\appendix

\section{\bf Recurrence Relations}

In this section we derive two basic recurrence relations which are useful in constructing ladder operators for the system (\ref{a5}). To begin with
we recall the Hermite differential equation,
\begin{eqnarray}
H^{''}_n - 2xH^{'}_n + 2nH_n = 0,
\label{ap1a}
\end{eqnarray}
and two of its associated recurrence relations, that is
\begin{eqnarray}
H^{'}_n = 2n H_{n-1},
\label{ap1b}
\end{eqnarray}
\begin{eqnarray}
2xH_n = H_{n+1}+2nH_{n-1},
\label{ap1c}
\end{eqnarray}
where prime denotes differentiation with respect to $x$. In Ref. \cite{Carinena}, the authors have
introduced a new family of polynomials ${\cal P}_n(x)$ (vide Eq. (8) in Ref. \cite{Carinena}) defined by
\begin{eqnarray}
{\cal P}_n(x) = H_n + 4nH_{n-2} + 4n(n-3)H_{n-4}, \quad n = 3,4,5,...
\label{ap2}
\end{eqnarray}
\begin{eqnarray}
{\cal P}^{'}_{n}(x) = 4n(1+2x^2)H_{n-3}
\label{ap8}
\end{eqnarray}
and
\begin{eqnarray}
\frac{{\cal P}_n(x)e^{-x^2}}{(1+2x^2)^2} = -2 \frac{d}{dx}\left[\frac{H_{n-3}}{1+2x^2}e^{-x^2}\right], \qquad n= 3,4,5,\ldots
\label{ap3}
\end{eqnarray}
Substituting Eqs. (\ref{ap1b}) and (\ref{ap1c}) in (\ref{ap3}) we get
\begin{eqnarray}
{\cal P}_{n}(x) = 2(1+2x^2)H_{n-2} + 8xH_{n-3}.
\label{ap7}
\end{eqnarray}
As our aim is to obtain a recurrence relation that connects ${\cal P}_n(x)$ with ${\cal P}^{'}_{n}(x)$ and ${\cal P}_{n-1}(x)$, we revoke Eq. (\ref{ap3}) in the form
\begin{eqnarray}
\frac{{\cal P}_{n-1}(x)e^{-x^2}}{(1+2x^2)^2} = -2 \frac{d}{dx}\left[\frac{H_{n-4}}{(1+2x^2)}e^{-x^2}\right], \qquad n= 4,5,6,\ldots
\label{ap4}
\end{eqnarray}
Substituting (\ref{ap1b}) in (\ref{ap4}), we find
\begin{eqnarray}
\frac{{\cal P}_{n-1}(x)e^{-x^2}}{(1+2x^2)^2} = -\frac{1}{(n-3)} \frac{d}{dx}\left[\frac{H^{'}_{n-3}}{(1+2x^2)}e^{-x^2}\right]
\qquad n= 4,5,6,\ldots
\label{ap5}
\end{eqnarray}
With the help of (\ref{ap1b}) and (\ref{ap1c}), Eq. (\ref{ap5}) can be brought to the form
\begin{eqnarray}
\fl \qquad (n-3){\cal P}_{n-1}(x) = [2n(1+2x^2)-(6+4x^2)]H_{n-3} - 4xH_{n-2}, \quad n = 4,5,6,...
\label{ap6}
\end{eqnarray}
Multiplying Eq. (\ref{ap7}) by $2x$ and Eq. (\ref{ap6}) by $(1+2x^2)$ and adding
the latter two equations and simplifying the resultant expression we arrive at
\begin{eqnarray}
\fl \qquad \frac{2x}{(1+2x^2)}{\cal P}_n(x) + (n-3){\cal P}_{n-1}(x) = 2(1+2x^2)\left[n-\frac{3+4x^4}{(1+2x^2)^2}\right] H_{n-3}.
\label{ap9}
\end{eqnarray}
Now multiplying Eq. (\ref{ap9}) by $2n$ and using (\ref{ap8}) in it,
one can obtain a recurrence relation which conects ${\cal P^{'}}_n(x)$ with
${\cal P}_{n}$ and ${\cal P}_{n-1}$, namely
\begin{eqnarray}
\fl \qquad \left[n-1+\frac{2(2x^2 - 1)}{(1+2x^2)^2}\right]{\cal P}_n^{'}(x)+n(x-\phi){\cal P}_n(x) = 2n(n-3){\cal P}_{n-1}(x),
\label{ap10}
\end{eqnarray}
where we have defined $\phi = x + \displaystyle \frac{4x}{1+2x^2}$.

Now we derive the second recurrence relation which connects ${\cal P}^{'}_n(x)$ with ${\cal P}_{n}(x)$ and ${\cal P}_{n+1}(x)$. To do so
we consider Eq. (\ref{ap3}) in the form
\begin{eqnarray}
\frac{{\cal P}_{n+1}(x)e^{-x^2}}{(1+2x^2)^2} = -2 \frac{d}{dx}\left[\frac{H_{n-2}}{(1+2x^2)}e^{-x^2}\right],\qquad n= 2,3,4,\ldots
\label{ap11}
\end{eqnarray}
Substituting Eqs. (\ref{ap1b}) and (\ref{ap1c}) in (\ref{ap11}) and simplifying it we obtain
\begin{eqnarray}
{\cal P}_{n+1}(x) =  - 4(n-2)(1+2x^2)H_{n-3}+2(x+\phi)(1+2x^2)H_{n-2}.
\label{ap12}
\end{eqnarray}
Multiplying Eq. (\ref{ap7}) by $(x+\phi)$ and subtracting it from (\ref{ap12}) and simplifying the resultant
expression, one obtains
\begin{eqnarray}
{\cal P}_{n+1}(x) - (x+\phi){\cal P}_n(x) = -4(1+2x^2)\left[n+\frac{2(2x^2-1)}{(1+2x^2)^2}\right] H_{n-3}.
\label{ap14}
\end{eqnarray}
Now one can replace Hermite polynomial $H_{n-3}$ in (\ref{ap14}) by ${\cal P^{'}}_n(x)$ (vide Eq. (\ref{ap8})).
As a result one gets
\begin{eqnarray}
-\left[n+\frac{2(2x^2-1)}{(1+2x^2)^2}\right]{\cal P}^{'}_n(x) + n(x+\phi){\cal P}_n(x) = n{\cal P}_{n+1}(x).
\label{ap15}
\end{eqnarray}
Recurrence relations (\ref{ap10}) and (\ref{ap15}) can be used to construct the ladder operators.


\section*{References}
\begin{thebibliography}{11}

\bibitem{Carinena}
Cari\~{n}ena J F, Perelomov A M, Ranada M F and Santander M 2008 \emph{J. Phys. A: Math. Theor.}
{\bf 41} 085301

\bibitem{Fellow}
Fellows J M and Smith R A 2009 \emph{J.Phys.A: Math.Theor.} {\bf 42} 335303

\bibitem{Sen}
Kraenkel R A and Senthilvelan M 2009 \emph{J. Phys. A: Math. Theor.} {\bf 42} 415303

\bibitem{Gbook}
Jean-Pierre Gazeau {\it Coherent states in Quantum Physics} (Wiley-VCH Verlag GmbH and Co, Weinheim, 2009)

\bibitem{chithiika}
Chithiika Ruby V and Senthilvelan M 2010 \emph{J. Math. Phys.} {\bf 51} 052106

\bibitem{Sesma}
Sesma J 2010 \emph{J. Phys. A: Math. Theor.} {\bf 43} 185303

\bibitem{Vbook}
Werner Vogel and Dirk-Gunnar Welsch {\it Quantum Optics} (Wiley-VCH Verlag GmbH and Co. Weinheim, 2006)

\bibitem{Nieto}
Nieto M M and Truax D R 1993 \emph{Phys. Rev. Lett.} {\bf 71} 2843

\bibitem{Manko1}
L\'{o}pez-Pe\~{n}a R, Man'ko V I, Marmo G, Sudarshan E C G and Zaccarrid F 2000 \emph{Journal of Russian Laser Research} {\bf 21} 4

\bibitem{Filho}
de Matos Filho R L and Vogel W 1996 \emph{Phys. Rev. A} {\bf 54} 4560

\bibitem{Manko}
Man'ko V I, Marmo G, Sudarshan E C G and Zaccarria F 1997 \emph{Phys. Scr.} {\bf 55} 528

\bibitem{Shan}
Shanta P, Chaturvedi S, Srinivasan V and Jagannathan R 1994 \emph{J. Phys. A: Math. Gen.} {\bf 27} 6433

\bibitem{Rokin}
Roknizadeh R and Tavassoly M K 2004 \emph{J.Phys.A: Math. Gen.} {\bf 37} 8111

\bibitem{Walls}
Walls D F and Zoller P 1981 \emph{Phys. Rev. Lett.} {\bf 47} 709;
Walls D F 1983 \emph{Nature} {\bf 306} 141

\bibitem{squee}
Roy B and Roy P 2000 \emph{J. Opt. B: Quantum Semiclass. Opt.} {\bf 2} 65;
Choquette J J, Cordes J G and Kiang D 2003 \emph{J. Opt. B: Quantum Semiclass. Opt.} {\bf 5} 56;
Wang J, Feng J, Liu T K and Zhan M S 2002 \emph{J. Phys. B: At. Mol. Opt. Phys.} {\bf 35} 2411;
L C Kwek and D Kiang 2003 \emph{J. Opt. B: Quantum Semiclass. Opt.} {\bf 5} 383

\bibitem{Paul}
Paul H 1982 \emph{Rev. Mod. Phys.} {\bf 54} 1061

\bibitem{Mah}
Mahran M H and Venkata Satyanarayana M 1986 \emph{Phys. Rev. A} {\bf 34} 640

\bibitem{Dodo1}
Dodonov V V 2002 \emph{J. Opt. B: Quantum Semiclass. Opt.} {\bf 4} R1 - R33

\bibitem{Vogel}
Kis Z, Vogel W and Davidovich L 2001 \emph{Phys. Rev. A.} {\bf 64} 033401

\bibitem{Manko2}
Man'ko V I, Marmo G and Zaccaria F 2010 \emph{Phys. Scr.} {\bf 81} 045004

\bibitem{Dong}
Shi-Hai Dong and Zhang-Qi Ma 2002 \emph{Am. J. Phys.} {\bf 70} 520

\bibitem{Ahmed}
Ahmed Jellal 2002 \emph{Mod. Phys. Lett. A} {\bf 17} 671

\bibitem{Barut}
Barut A O and Girardello L 1971 \emph{Commun. Math.Phys.} {\bf 21} 41

\bibitem{Gazeau}
Gazeau J P and Klauder J R 1999 \emph{J.Phys.A: Math. Gen.} {\bf 32} 123

\bibitem{Klauder}
Klauder J R and Skagerstam B S Coherent States: Applications in Physics and Mathematical Physics
 (Singapore: World Scientific, 1985); Perelomov A M  Generalized Coherent States
and Their Applications (Berlin: Springer, 1986)


\bibitem{Glau}
Glauber R J 1963 \emph{Phys. Rev.} {\bf 131} 2766

\bibitem{Schr}
Schr\"{o}dinger E Sitzungsber. Preuss. Akad. Wiss, Phys-Math. Klasse 19 (Berlin: Springer,1930) 296

\bibitem{Robert}
Robertson H P 1930 \emph{Phys. Rev.} {\bf 35} 667A;
Robertson H P 1934 \emph{Phys. Rev.} {\bf 46} 794

\bibitem{Arag}
Aragone C, Guerri G, Salamo S and Tani J L 1974 \emph{J. Phys. A: Math. Nucl. Gen.} {\bf 7} L149

\bibitem{Dodo}
Dodonov V V, Kurmyshev E V and Mank'o V I 1980 \emph{Phys. Lett. A} {\bf 79} 150


\bibitem{Kinani}
El Kinani A H and Daoud M 2002 \emph{J. Math. Phys.} {\bf 43} 714


\bibitem{Man}
Mancini S 1997 \emph{Phys. Lett. A} 233 291

\bibitem{Siva}
Sivakumar S 1998 \emph{Phys. Lett. A } {\bf 250} 257

\bibitem{Koc}
Koc R, Koca M and Sahinoglu G 2005 \emph{Eur. Phys. J. B} {\bf 48} 583

\bibitem{Miller}
Gossard A C, Miller R C and Wiegmann W 1986 {\emph Surf. Sci.} {\bf 174} 131




\bibitem{Klauder2}
Klauder J R, Penson K A and Sixdeniers J M 2001 \emph{Phys. Rev. A} {\bf 64} 013817

\bibitem{Chat}
Chaturvedi S 1996 \emph{Mod. Phys. Lett. A} {\bf 11} 2805

\bibitem{book}
Gradshteyn I S and Ryzik I M {\it Table of Integrals, Series and Products} (Academic Press, Inc.  1980)




\bibitem{Mandel}
Mandel L 1979 \emph{Opt. Lett.} {\bf 4} 205

\bibitem{Ant}
Antoine J P, Gazeau J P, Monceau P,  Klauder J R and Penson K A 2001 {\emph J. Math. Phys.} {\bf 42} 2349

\bibitem{Tava}
Tavassoly M K 2006 {\emph J. Phys. A: Math. Gen.} {\bf 39} 11583

\bibitem{Shree}
Shreecharan T and Shiv Chaitanya K V S 2010 {\emph Aspects of coherent states of nonlinear algebras} arxiv:1005.5607v1

\bibitem{pdms}
Plastino A R, Rigo A, Casas M, Gracias F and Plastino A 1999 \emph{Phys. Rev. A} {\bf 60} 4318


\end{thebibliography}
\end{document}